\documentclass[fleqn,twoside]{article}
\usepackage{espcrc2}
\usepackage{amssymb}

\usepackage{epsfig}
\usepackage{graphicx}
\usepackage[figuresright]{rotating}

\def\kms {\hbox{${\rm km\ s}^{-1}$}}
\def\scm  {$\hbox{{\rm cm}}^{-2}$}    
\newcommand{\NHI}{N_{\rm HI}}
\newcommand{\dV}{\mathrm{d} V}
\newcommand{\ts}{T_{\rm s}}
\newcommand{\tk}{T_{\rm k}}

\def\lapp{\ifmmode\stackrel{<}{_{\sim}}\else$\stackrel{<}{_{\sim}}$\fi}
\def\gapp{\ifmmode\stackrel{>}{_{\sim}}\else$\stackrel{>}{_{\sim}}$\fi}

\newcommand{\AmS}{{\protect\the\textfont2
  A\kern-.1667em\lower.5ex\hbox{M}\kern-.125emS}}

\title{21 cm Absorption Studies with the Square Kilometer Array}

\author{
N. Kanekar\address{Kapteyn Institute, University of  Groningen, The Netherlands},
F. H. Briggs\address{RSAA, The Australian National University,
Mount Stromlo Observatory, Australia \\ $^{\rm b}$Australia Telescope National
Facility, Sydney, Australia}}

\begin{document}

\begin{abstract}

HI 21~cm absorption spectroscopy provides an excellent probe of the
neutral gas content of absorbing galaxies, yielding information on
their kinematics, mass, physical size and ISM conditions. The high
sensitivity, unrivaled frequency coverage and RFI suppression
techniques of the SKA will enable it to use HI absorption to study the
ISM of high column density intervening systems along thousands of
lines of sight out to high redshifts. Blind SKA 21~cm surveys will
yield large, unbiased absorber samples, tracing the evolution of
normal galaxies and active galactic nuclei from $z \gtrsim 6$ to the
present epoch. It will thus be possible to directly measure the
physical size and mass of typical galaxies as a function of redshift
and, hence, to test hierarchical models of structure formation.

\end{abstract}

\maketitle

\section{INTRODUCTION}
\label{sec:intro}

The epoch at which objects of a particular mass scale form is a
critical test of cosmological models. Theoretically favored structure
formation scenarios\footnote{Including the currently standard
$\Lambda$-CDM ``concordance cosmology'', with $\Omega_m=0.27$,
$\Omega_{\Lambda}=0.73$, $H_0=71$~\kms~Mpc$^{-1}$; this model will be
used for numerical estimates in the current article.}  predict that
large galaxies are formed in hierarchical fashion, with smaller cold
dark matter (CDM) halos merging to form larger ones, followed by the
dissipative re-collapse of baryons into the resulting deep
gravitational potentials (e.g. \cite{white91}). Such CDM-based
hierarchical merger models have the generic feature that big galaxies
are formed at late times ($z \lesssim 1$), in order to remain
consistent with the observed anisotropies in the microwave
background. The period between redshifts $z \approx 5$ and $z \approx
1$ is thus expected to be one of vigorous assembly of galaxies.

Measuring the typical size and mass of galaxies as a function of
redshift allows us to directly test the hierarchical merger
paradigm. Unfortunately, this has been difficult to carry out in
practice. While both the number density of luminous quasars and the
star formation rate (SFR) density in Lyman-break galaxies appear to
peak at $z \approx 2$ (e.g. \cite{madau96,schmidt95,steidel99}), these
observations are strongly biased toward the bright end of the
luminosity function and may not probe the behavior in ``typical''
systems. Further, the stellar mass detected in Lyman-break galaxies at
these redshifts is but a fraction of the neutral gas mass in all
galaxies, implying that the observations are not sensitive to the bulk
of the baryons. 


One of the SKA goals is the direct monitoring of the HI content of
collapsed objects, from the epoch of formation right through to the
present day (see contribution by van der Hulst et al., this Volume). In fact,
the square kilometer of aperture is a consequence of the interest in
mapping gas-rich galaxies out to redshifts as high as $z\sim 1$.  This
stems from a natural desire to image the high-$z$ galaxies in the same
fashion as done by the VLA and WSRT for systems at $z<0.05$.  Alas,
the frontier of galaxy evolution studies has moved to higher
redshifts, where even the SKA will have trouble detecting massive
galaxies and will not make maps with significant structural and
kinematic information.

HI~21~cm absorption measurements toward radio-loud background sources
(either quasars or radio galaxies) can provide interesting information
on the structure and physical conditions in high redshift galaxies
(e.g. \cite{kanekar03} and references therein). To date, a number of
technical factors (e.g. frequency coverage, spectrometer
bandwidth/resolution, RFI environment, adequate collecting area, etc)
have combined to hinder such studies with current radio telescopes. We
describe in this chapter the potential of the SKA to improve our
understanding of galaxy evolution through absorption studies in the
HI~21~cm line.

\section{DAMPED LYMAN-$\alpha$ SYSTEMS}
\label{sec:dla}

\begin{figure}[t]
\epsfig{file=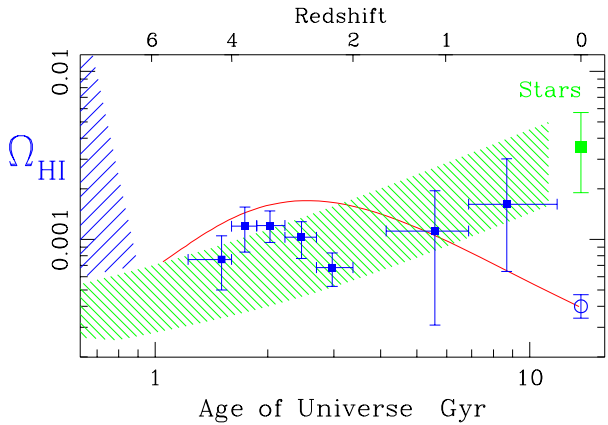,width=3.3in}
\caption{Neutral gas content of the Universe as a function of time,
expressed as a fraction of the critical density at the present epoch.
The filled squares for $5>z>2$ are from ``blind'' optical DLA surveys
\cite{prochaska04}, the filled squares for $1.6>z>0$, from MgII-selected
DLA surveys \cite{nestor03} and the  open circle at $z=0$, from HI emission 
surveys \cite{zwaan03}. The light shaded  region indicates the rising trend
of stellar mass with time, leading to the filled square for the present 
stellar mass in galaxies \cite{fukugita98}.  
The solid curve for the evolution of HI
density with time indicates the spirit of the dust-corrected models of 
\cite{pei99}. The darker shading to the left at $z>6$ reflects the rise in
baryon neutral fraction during the Epoch of Reionization \cite{becker01,fan02}.
}
\label{fig:omega}
\end{figure}

Absorption line studies toward bright background sources provide a
powerful observational probe of galaxy formation and evolution. While
emission studies of a flux-limited, high redshift sample are usually
dominated by the brightest (often atypical) sources, absorption
samples are likely to be more representative of ``normal'' galaxies at
a given redshift, since the most common objects that host HI clouds
will provide the largest cross section for chance intervention toward
high $z$ radio sources. The high HI~column density absorbers, the
damped Lyman-$\alpha$ systems (DLAs, with $\NHI \gtrsim 10^{20}$~\scm)
are of particular interest as these are the largest repository of
neutral gas at high redshifts and hence believed to be the precursors
of present-day galaxies (e.g. \cite{wolfe86}). As indicated in
Figure~\ref{fig:omega}, the HI mass density in DLAs at $z \sim 3$ is a
factor $\sim 4$ larger than that observed today, but comparable to the
stellar mass density in luminous galaxies at $z \sim 0$
(e.g. \cite{storrie00}). This is consistent with a picture of gas
depletion with time due to star formation, exchanging the neutral gas
density for mass locked in populations of long-lived stars.

Understanding the evolution of a ``typical'' DLA with redshift is
 critical to understanding normal galaxy evolution. Specifically, one
might directly test structure formation models by measuring the
typical sizes, structures and dynamical masses of an unbiased sample
of DLAs as a function of redshift.  Such a sample could also be used
to understand the evolution of physical conditions in the interstellar
medium of normal galaxies, the onset of star formation, etc.

Despite twenty five years of detailed study, the typical nature of
high $z$ DLAs remains a controversial issue, with models ranging from
large, rapidly rotating massive disks (e.g. \cite{prochaska97}) to
small, merging sub-galactic systems
(e.g. \cite{haehnelt98}). Similarly, physical conditions in the ISM of
the absorbers are also the subject of much debate, with some authors
claiming evidence that most of the HI~is in a cold phase
\cite{wolfe03} and others arguing for predominantly warm gas
\cite{kanekar03,petitjean00}.  Studies of DLA chemical abundances have
also found only weak evidence for evolution
\cite{pettini99,prochaska03}, with low metallicities ([Zn/H] $\lesssim
-1$) typical even at low redshifts. This is somewhat surprising if the
absorber population evolves into the population of large galaxies of
solar metal abundance that dominate the absorption cross-section in
the Universe today \cite{rao93,zwaan99,zwaan02}.

There are a number of reasons for the general contention surrounding
the nature of damped absorbers. Present DLA samples contain two
obvious biases, which complicates their use in studying the evolution
of the average galaxy population.  First, it has only been possible to
carry out extensive spectroscopic surveys for DLAs at redshifts
$z_{\rm abs} \gtrsim 1.7$, as it is here that the Lyman-$\alpha$ line
is observable with ground-based telescopes; current DLA samples are
hence strongly biased toward high redshifts, presenting a snapshot of
the Universe during the time window at 10~--~30\% of its present age.
The redshift range $0 < z \lesssim 1.7$ comprises 70\% of the age of
the Universe, covering the stages where proto-galactic systems evolve
to present-day galaxies in hierarchical merger scenarios. Further, the
high $z$ DLA samples are almost entirely drawn from optically selected
surveys (e.g. \cite{wolfe86,storrie00}; but see \cite{ellison01}) and
may be biased against systems with a high metallicity (i.e. a high
dust content; see, for example, \cite{fall93}). Piecing together the
puzzle critically requires an unbiased sample of DLAs with a uniform
sampling in redshift; as will be seen later, this can be achieved
through absorption surveys in the HI~21~cm line.

In addition to current biases, it has so far proved very difficult to
distinguish between different models by direct observations of DLAs at
high redshift. Direct imaging is restricted by large observing time
requirements at all wavebands; optical imaging studies are further
complicated due to problems with subtracting out the point spread
function of the background QSO. Similarly, it is not possible to
determine the transverse size of the absorbers using optical or
ultraviolet absorption lines as the background AGN source is
unresolved at these wavelengths; even cases of multiply lensed images
provide very few lines of sight through the intervening galaxy.  Radio
absorption studies provide perhaps the only hope in this regard as it
is only in this waveband that some sources have extended background
continuum sources, against which one might determine the transverse
size and kinematics of the absorbing galaxy. Of course, the vast
quantities of neutral hydrogen at high redshifts make the HI~21~cm
line the obvious transition for this purpose.

\section{21~CM ABSORPTION STUDIES }
\label{sec:21cm}

\begin{figure*}[ht]
\epsfig{file=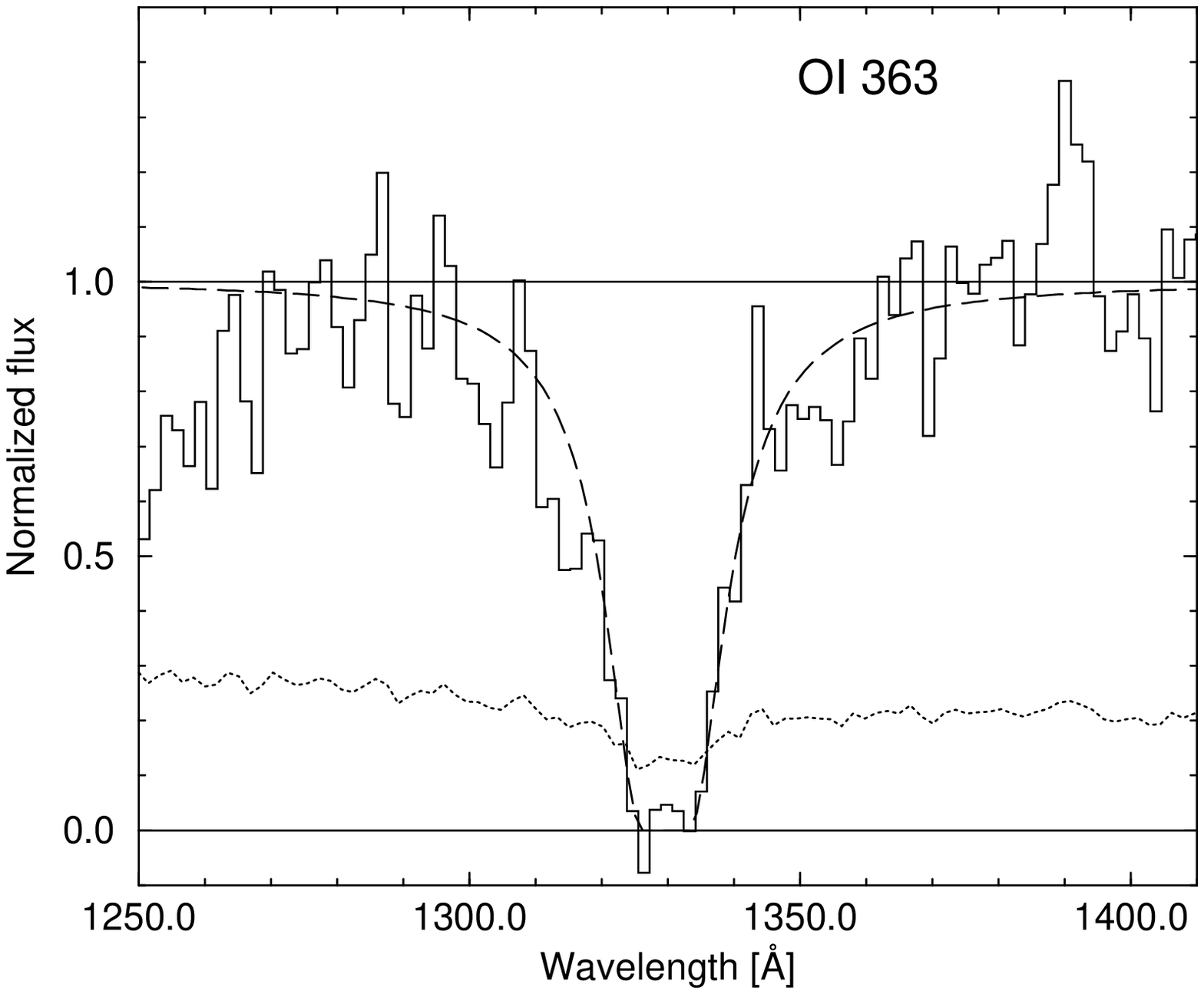,width=3.1in, height=3.1in}
\epsfig{file=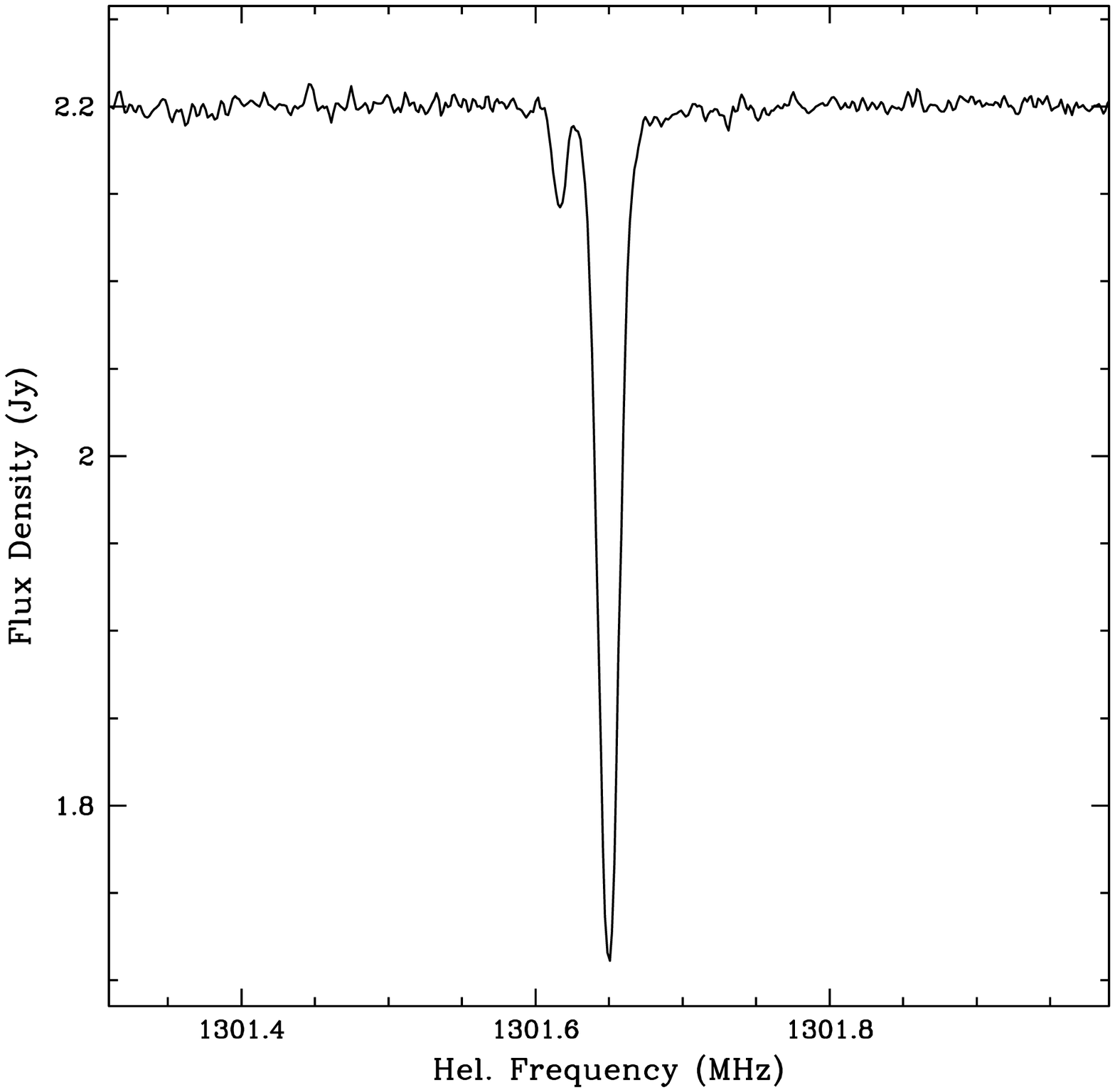,width=3.3in,height=3.3in}
\caption{[A] Left panel : The $z = 0.0912$ damped Lyman-$\alpha$ line 
	toward QSO~0738+313, obtained with HST-FOS \cite{rao00}.
	[B] Right panel : The 21~cm profile of the above absorber, obtained 
	with Arecibo \cite{lane00}.
}
\label{fig:oi363}
\end{figure*}

The HI column density $\NHI$ of a homogeneous cloud in
thermal equilibrium is related to its 21~cm optical depth 
$\tau_v $ and spin temperature $\ts$ by the expression
\begin{equation}
\label{eqn:21cm}
\NHI = 1.823 \times 10^{18} \: \ts  \int \tau_v\: \dV \:\: ,
\end{equation}
where  $\NHI$ is in cm$^{-2}$,
$\ts$ in K and $\dV$ in km~s$^{-1}$. 
In the optically thin limit, this expression relates $\NHI$
to the emission brightness temperature $T_B$ as
$\NHI = 1.823 \times 10^{18} \:   \int T_B(V)\: \dV$.
When optical depth is computed from the ratio of flux density in the
absorption line depth $\Delta S_{\nu}$ to the integrated
source continuum $S_{\nu}$,
the assumption is often made that a layer of uniform spin temperature and
HI column density covers a fraction $f$ of the background source, leading
to $\tau_{\nu}=-\ln (1-\Delta S_{\nu}/fS_{\nu})$

The relatively weak strength of the 21~cm transition implies that 21~cm
absorption is only detectable in high column density gas. This can be
clearly seen from Equation~(\ref{eqn:21cm}); typical 21~cm equivalent
widths\footnote{Note that the equivalent width is defined here as
$W_V \equiv \int \tau\: \dV$ (e.g. \cite{kanekar03}), unlike the usual
definition ${W_{\lambda}\equiv\int (1-e^{-\tau_{\lambda}})\:d\lambda}$,
used when $\tau_{\lambda} \gtrsim 1$} are $W_V \sim 1$~km~s$^{-1}$ while spin
temperatures are  usually $\gtrsim 100$~K. These parameters lead to typical HI
column densities $\NHI \gtrsim 10^{20}$~\scm, precisely those that
would give rise to damping wings in the Lyman-$\alpha$ line. 21~cm absorption
surveys are thus well-matched toward the construction of DLA samples. Further,
such surveys contain no ``intrinsic'' redshift bias, such as that arising from
the ultraviolet cut-off in the atmosphere. And, of course, flux-limited radio
samples are unaffected by dust. Blind surveys in the 21~cm line could be
used to obtain DLA samples unbiased by dust extinction and redshift coverage.
On the other hand, the spin temperature dependence inflicts a bias toward
cold gas on blind surveys, and the allocation of radio frequency bands to
communications and navigation services excludes those bands from blind
surveys for observatories located in industrialized regions of the world.

Figure~\ref{fig:oi363} shows a comparison between the damped
Lyman-$\alpha$ and 21~cm spectra for the $z \sim 0.0912$ DLA toward
QSO~0738+313 \cite{rao00,lane00}. The Lyman-$\alpha$ line is highly
saturated, allowing the DLA line to provide no kinematic
information on the absorbing galaxy; the less saturated metal
lines in these systems are often diagnostics of complex kinematic
structure \cite{prochaska97}.  Unlike the 21cm absorption, 
the DLA line is
insensitive to the temperature of the absorbing gas. 
While the 21~cm absorption in this $z \sim 0.0912$ 
absorber is stronger than usual in redshifted
systems, the peak optical depth is still only $\sim 0.24$, implying
that the profile might be used to glean information on the kinematics
of the absorber. In fact, the 21~cm absorption profile in this
absorption against QSO~0738+313 could be decomposed
into three components, two arising from absorption in cold ($\sim
100$~K) gas and the third in the warm phase, at a kinetic temperature
of $\sim 5500$~K \cite{lane00}. Thus, while the damped Lyman-$\alpha$
line directly provides the HI column density along the line of sight,
21~cm studies are extremely useful in detailed studies of both
kinematics and physical conditions of gas in galaxies.

\section{HI KINEMATICS : SPATIAL MAPPING OF 21~CM ABSORPTION}
\label{sec:map}

\begin{figure*}[t]
\epsfig{file=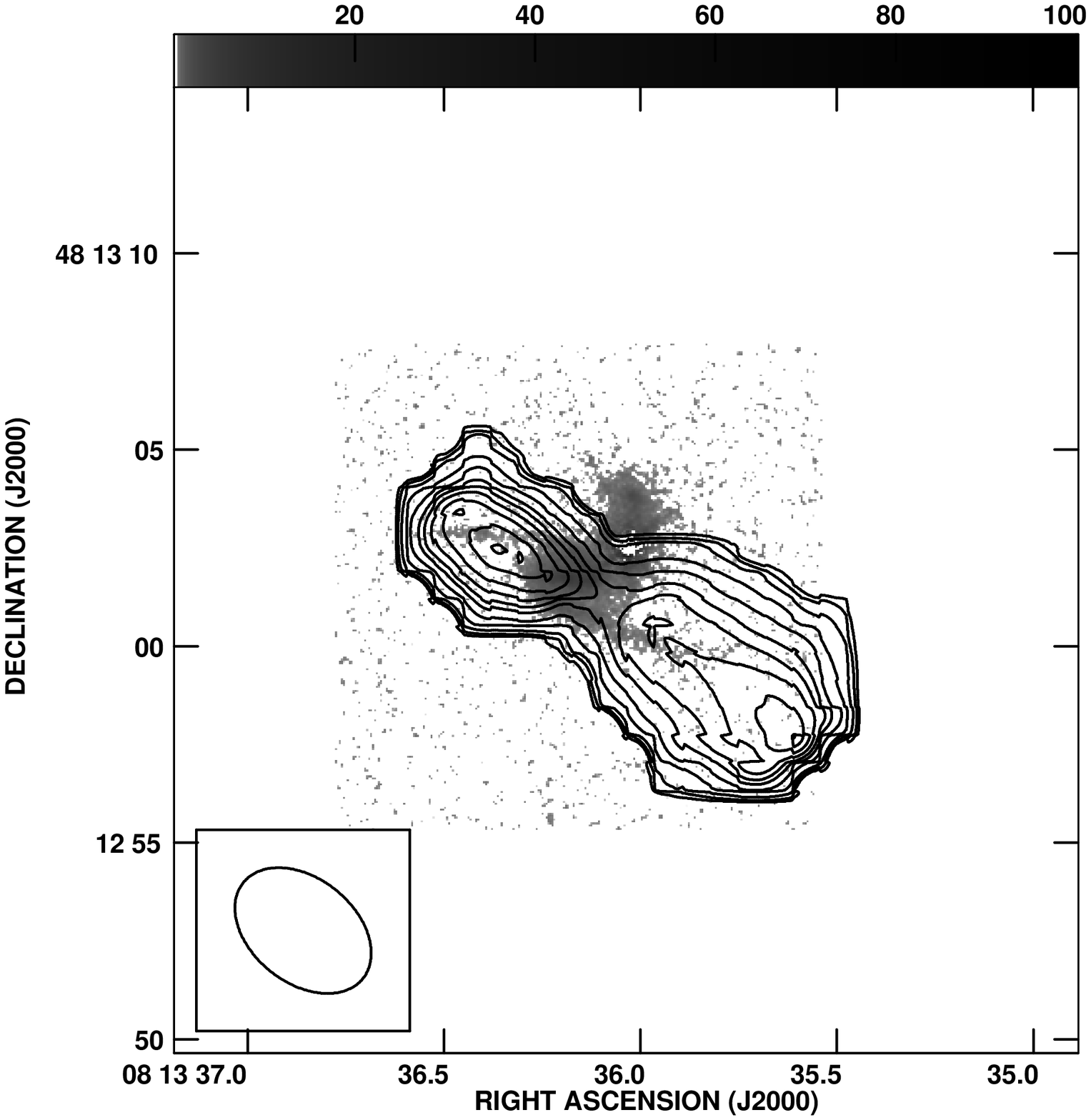,width=3.0in, height=3.0in}
\epsfig{file=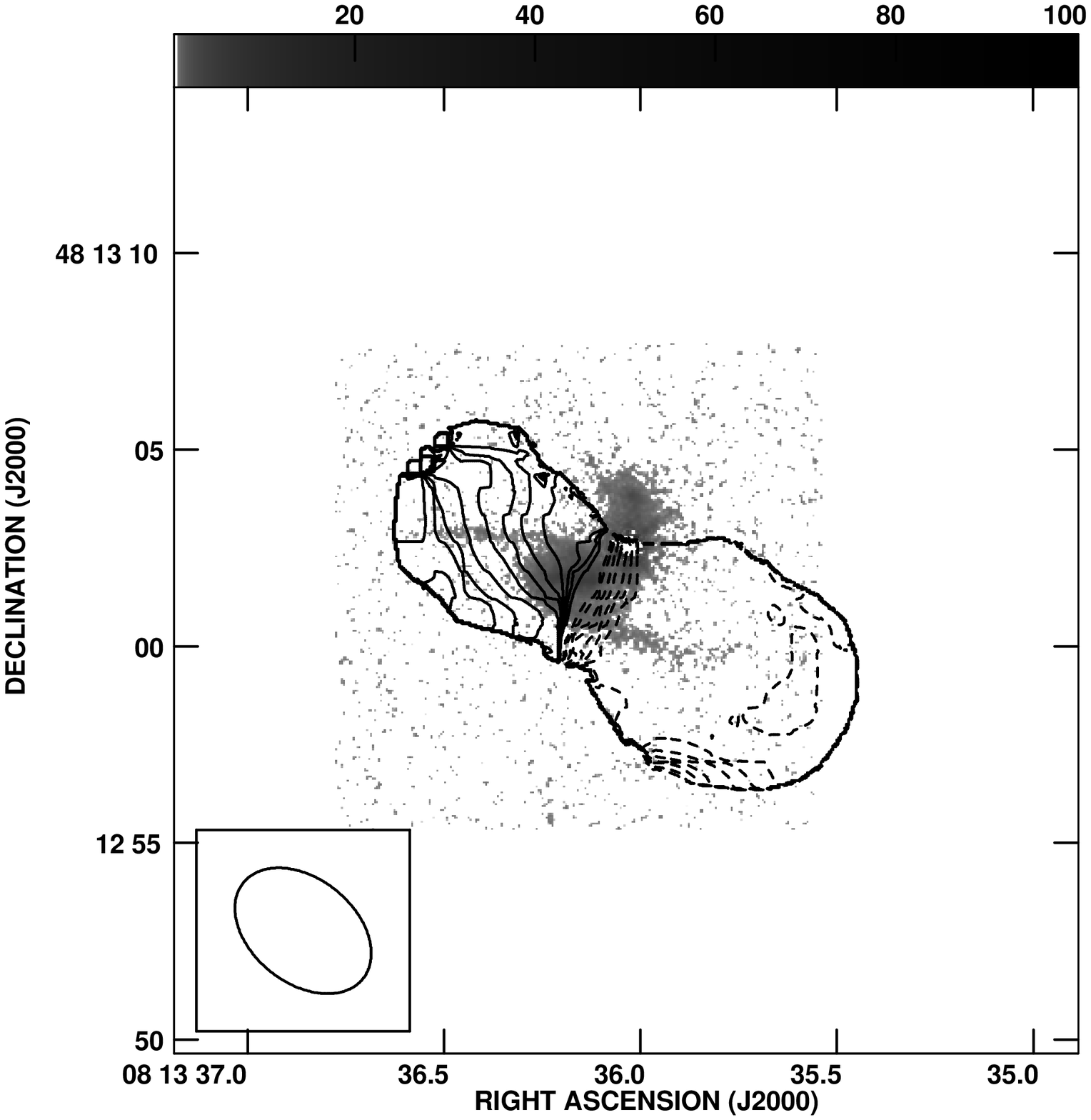,width=3.0in,height=3.0in}
\caption{[A] Left panel : The integrated 21~cm optical depth toward
        3C196 (contours; \cite{kanekar04b}), overlaid on an HST image 
	(greyscale; \cite{ridgway97}).
	[B] Right panel : The integrated 21~cm absorption velocity field
	toward 3C196 (contours; \cite{kanekar04b}), overlaid on an HST image 
	(greyscale; \cite{ridgway97}).}
\label{fig:3c196}
\end{figure*}

The goals of mapping the gas components of galaxies at high redshift
are: (1) determine the physical sizes of the galaxies, (2) measure the
kinematics of the system, (3) assess whether the system is disk-like
or a recent, turbulent remnant of formation, and (4) if disk-like,
compute the effective dynamical mass for use in testing theories for
the assembly of galaxies. The most direct way of doing this is through
21~cm {\it emission} observations; unfortunately, the sensitivity of
current-day radio telescopes limits such studies to local objects. In
fact, we will see in Section~\ref{sec:ska} that prohibitively large
amounts of time would be needed even with the SKA to image large
samples of high $z$ galaxies in the 21~cm line.

Constraints on the size and gas kinematics of DLAs can also be
obtained from high angular resolution 21~cm {\it absorption} studies,
provided the background radio continuum is extended on scales larger
than the telescope resolution element \cite{kanekar04b}. Of course,
there are differences between kinematic studies in absorption and in
emission.  Spatially resolved emission studies directly provide the HI
mass of the galaxy\footnote{Assuming that the emission is optically
thin.}, its physical extent, velocity field and dark matter
potential. On the other hand, less information is available in
absorption since the observations are only sensitive to the kinematics
of gas lying in front of the background radio continuum
(e.g. \cite{briggs00}). Further, mapping the absorption does not
measure the HI mass directly but only the optical depth; one must
assume a spin temperature to obtain an estimate of
the HI mass of the absorber\footnote{$\ts \sim
100$~K provides a fairly good lower limit to the mass.}. Absorption
and emission studies are also sensitive to slightly different velocity
fields, weighted by optical depth against the structure in the
background continuum source in the former case and by column density
in the latter. Both provide information on whether the velocity field
is quiescent (i.e. well-ordered) or whether disruptive events
(e.g. mergers) have occurred in the recent past. Given sufficient
spatial resolution and sensitivity, it is possible to model the
absorption velocity field to obtain the dynamical mass.  In fact, even
in the absence of high spatial resolution 21~cm line observations,
spectral modeling in combination with knowledge of the background
radio structure can constrain properties of the absorbing galaxy
\cite{briggs01,kanekar03b}. Finally, absorption mapping provides a
lower limit to the physical extent of the galaxy.  The full
application of absorption studies will generally require 
statistical analysis of a large number of systems.

Spatial mapping of 21~cm absorption requires the identification of
DLAs toward extended radio sources such as radio galaxies. The median
size of high $z$ radio galaxies ($\sim 50$~kpc) is well matched toward
testing whether large gaseous disks are common at high redshift. Radio
galaxies usually have weak optical counterparts, implying that it is
difficult to find such systems through optical spectroscopic surveys,
since the latter target optically bright QSOs, which tend to have
compact radio structure. On the other hand, surveys selecting DLA
absorbers through 21~cm absorption toward radio galaxies would also
provide interesting objects for follow-up optical imaging. PSF
subtraction should not be an issue here as there would be little
bright optical continuum; this would aid the optical identification of
the intervening galaxy.

A number of VLBI 21~cm observations have studied absorption against 
compact radio components of the background QSO, at high spatial resolution
(e.g. \cite{wolfe76,lane00,briggs89}). However, at present, the paucity of 
absorbers toward radio galaxies and poor frequency coverage of radio 
interferometers has meant that mapping studies sensitive to galactic disk 
scales of $\sim 10$~kpc have only been possible in two DLAs, at $z \sim 0.4$.
Figs.~\ref{fig:3c196}[A] and [B] show the zeroth and first velocity moment
of the integrated 21~cm absorption toward 3C196, obtained with the GMRT
\cite{kanekar04b}.  The two main components of the 21~cm  profile were
found to arise from absorption in two arms of a large barred spiral,
against the south-west hotspot and the eastern lobe of 3C196, as had been
deduced by \cite{briggs01}. HI absorption was detected out to a radius of
$\sim 70$~kpc, far beyond the extent of the optical galaxy. The authors
have also resolved the 21~cm absorption from the $z \sim 0.395$ DLA toward
PKS~1229$-$021, showing that the absorber has a physical size larger than
$\sim 30$~kpc, again a result that had been earlier inferred from modeling
the unresolved 21~cm spectrum obtained at the WSRT \cite{briggs99}.

It should be pointed out that current telescopes are very unlikely to
make significant advances in mapping 21~cm absorption at redshifts
higher than $z_{\rm abs} \sim 0.5$. Sensitive interferometers such as
the GMRT and the VLA do not have sufficiently long baselines to map
the absorption; for example, the GMRT has a spatial resolution of
$\sim 25$~kpc at 610~MHz, i.e. at $z_{\rm abs} \sim 1.3$. At the other
extreme, very long baseline interferometers like the VLBA resolve out most of
the extended background emission, due to the absence of short
baselines. Progress in this field requires interferometers with
wideband frequency coverage and modest angular resolution ($0.1''$) to
resolve $\lesssim 1$~kpc at $z \sim 1 - 5$ (corresponding to baseline
lengths $B \gtrsim 600 (1 + z) $~km).  Mapping of complex sources
would also require good U-V coverage and a high sensitivity. The large
antennas of the EVN+Merlin would be a good match to this problem in
the near term but for the high level of background radio communication
signals throughout Europe, which lessen the feasibility for equipping
the arrays with appropriate receivers. Some progress in mapping
absorption against background sources of simple structure might be
made through continuum mapping in standard radio astronomy bands that
bracket the absorption frequency, followed by line observations with a
few well chosen baselines and model fitting to tie the HI absorption
features to specific lines of sight.

\section{ISM CONDITIONS : THE SPIN TEMPERATURE}
\label{sec:tspin}

\begin{figure}[h]
\centering
\epsfig{file=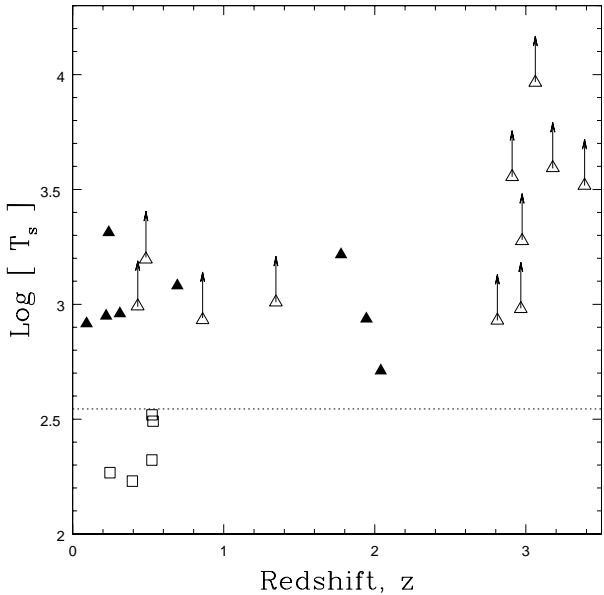,height=3.3in,width=3.3in}
\caption{The spin temperature (Log[$T_{\rm s}$]) as a function of redshift 
for the 24 DLAs of the present sample \cite{kanekar03}. Objects identified 
with spiral galaxies are plotted using squares; other detections are shown as 
filled triangles, with non-detections shown as open triangles, with arrows. } 
\label{fig:tvsz}
\end{figure}

If the HI column density toward a radio-loud QSO can be obtained from
independent observations (e.g. from the damped Lyman-$\alpha$ profile,
soft X-ray absorption, etc), a measurement of the 21~cm optical depth can
be used in equation~(\ref{eqn:21cm}) to estimate the spin temperature.
In contrast to mapping studies, these experiments are most easily 
accomplished with compact background sources,
so that the line of sight toward the optical or ultraviolet source
is the same as that toward the radio continuum. For more complex sources,
VLBI mapping is necessary to isolate only those 21cm absorbing clouds that
contribute to absorption against the QSO nucleus.

In the Galactic interstellar medium, the spin temperature $\ts$ of
cold, dense HI clouds (comprising the ``cold neutral medium''; CNM) is
simply the kinetic temperature $\tk$ of the cloud
\cite{field58,kulkarni88}; on the other hand, $\ts \le \tk$ for the
lower density, ``warm neutral medium'' (WNM) \cite{liszt01}.  For
multi-phase structure along a line of sight, the net $\ts$ is the
column-density-weighted harmonic mean of the temperatures of different
phases (e.g. \cite{kulkarni88}), and $\ts$ is biased toward the cold
gas. For example, assuming typical temperatures of $\sim 100$~K and
$\sim 8000$~K for the cold and warm phases, a line of sight with half
the gas in each phase would have $\ts \sim 200$~K, while one with 90\%
of the gas in the WNM and only 10\% in the CNM would have $ \ts \sim
900$~K.  The spin temperature thus provides information on the
fraction of neutral gas at different temperatures, that might be used
to study the evolution of the ISM with redshift.

Even more interesting, however, $\ts$ may serve as an indicator of
galaxy type. This is because the global fractions of cold and warm
neutral gas in a galaxy are likely to be primarily determined by its
central pressure and average metallicity \cite{wolfire95}. Large
spiral galaxies like the Milky Way are expected to have relatively
large amounts of CNM, as such systems have both a high central
pressure (due to their large mass) and a high metallicity; the Milky
Way has roughly equal amounts of warm and cold HI. On the other hand,
the low metallicities and pressures in smaller systems like dwarf
galaxies are not conducive to the formation of the cold phase. This
occurs because the ISM in small galaxies cools inefficiently, due to
the lack of metals (and hence a lack of radiation pathways); such
systems are hence expected to have a substantially higher HI fraction
in the warm phase. An average line of sight through a mature large
spiral should thus have a systematically lower spin temperature than
an average line of sight through a low-metallicity, dwarf galaxy.

Typical lines of sight in the Milky Way and local spirals like
Andromeda have $\ts \lesssim 200$~K (e.g. \cite{braun92}). In 
contrast, two decades of observations have found that the majority of DLAs have
significantly higher spin temperatures, $\ts \gtrsim 700$~K
(e.g. \cite{wolfe79,wolfe81,wolfe85,taramopoulos94,carilli96,briggs97,lane98,chengalur98,chengalur00,kanekar01,kanekar03}).
Figure~\ref{fig:tvsz} shows a plot of the measured spin temperatures
as a function of redshift for the 24 DLAs of the present 21~cm
absorption sample \cite{kanekar03}. Both low and high $\ts$ values are
found at low redshift $z < 1$, while all high redshift ($z \gtrsim 3$)
DLAs have high $\ts$. Present results thus suggest that high $z$ DLAs
have fairly small CNM fractions, with most of the gas in the warm
phase. Figure~\ref{fig:tvsz} also hints at evolution in the relative
fraction of the cold and warm phases from high redshifts to today.

The key to understanding the evolution in $\ts$ is
the anti-correlation found between $\ts$ and metallicity
[Zn/H] in a sample of 15 DLAs drawn from all redshifts
\cite{kanekar04}; this finding implicates gas metallicity in
determining the temperature of the neutral ISM. At high redshift, there
is a larger fraction of primitive, low-enrichment ISM gas in the
cross section provided by the random intervening clouds.

In the low redshift range of the sample plotted in
Figure~\ref{fig:tvsz}, all five low $z$, low $\ts$ DLAs have been
identified with luminous spiral galaxies \cite{lebrun97,rao03}. In
contrast, the low $z$ DLAs with $\ts \gtrsim 700$~K have been found to
be associated with dwarf or low surface brightness (LSB) systems
(e.g. \cite{rao03,cohen01}). The current low $z$ sample thus suggests
that $\ts$ can indeed ``distinguish'' between intervening dwarfs and large
spirals for galaxies in the nearby universe. This correlation relies on 
the primitive nature of the dwarf irregular and LSB galaxies so that
their level of metal enrichment lies below that of the more evolved
large spirals, and hence their cooling is less efficient, and their
observed spin temperature is higher.

While it may be tempting to argue that high redshift absorbers with
high spin temperatures are small galaxies, the only clear physical
evidence on their nature is that their ISM gas is primitive material
of low-metallicity.
Determination of physical sizes and masses of the absorbers awaits
measurement by the next generation of radio telescopes.

\section{ASSOCIATED 21~CM ABSORBERS }
\label{sec:assoc}

Blind 21~cm surveys will also yield large samples of {\it associated}
absorption systems, probes of the local environments of radio galaxies
and quasars
(e.g. \cite{vangorkom89,conway95,morganti01,vermeulen03,pihlstrom03}). This
will allow detailed studies of the kinematics and distribution of gas
close to the AGN, important for understanding the physics of AGN
activity, especially since this activity may well be fueled (and
perhaps even triggered) by the neutral gas
(e.g. \cite{vangorkom89}). Such observations allow direct tests of the
unification scheme for radio galaxies and quasars
(e.g. \cite{barthel89}).  For example, such schemes predict that the
line of sight to broad line radio galaxies is normal to the torus (and
the thick disk), while that toward narrow line radio galaxies lies
close to the plane of the torus. Associated 21cm absorption should
thus be systematically more common in the latter class of systems as
the nucleus in broad line systems is not expected to be obscured by
the neutral gas.  Preliminary evidence for this effect has already
been found (e.g. \cite{morganti01}), but the number of systems in the
sample is quite small and restricted to fairly low redshifts ($z
\lesssim 0.2$).  Similarly, van Gorkom et al.  \cite{vangorkom89} find
a clear preponderance of infall in local radio galaxies while
Vermeulen et al. \cite{vermeulen03} show tantalizing evidence that
outflows are more common than infall at intermediate redshifts ($0.2
\lesssim z \lesssim 0.8$). While the two results hint at evolution in
the nuclear environment, small number statistics again make it
difficult to draw firm conclusions. The SKA 21~cm surveys (see also the 
contribution of Jarvis \& Rawlings, this Volume) will provide 
homogeneous samples of associated absorbers out to at least $z \sim
6$, allowing us to trace in detail any evolution in nuclear
environments (e.g. disk size, the existence of outflows and infall,
etc) from high redshifts to the present epoch.

\section{PROSPECTS WITH THE SKA}
\label{sec:ska}

An important step to using DLAs as tracers of galaxy evolution is
setting up an unbiased absorber sample; this can be done through blind
21~cm absorption surveys toward radio-loud sources.
The weakness of the 21~cm transition implies that high sensitivity is
needed for such surveys, especially since they must perforce be
carried out at fairly high spectral resolution (21~cm lines can be
fairly narrow, with FWHM~$\gtrsim 5$~\kms; e.g. \cite{chengalur99}).
Further, for a cosmologically distant absorber, the 21~cm line
redshifts out of the ``protected'' 1420~MHz radio band into frequency
ranges allocated to communication and navigation services; carrying
out astrophysical observations at such frequencies is a serious
challenge. Of course, wideband frequency coverage is a pre-requisite
to carrying out the survey observations. And, finally, mapping
absorbers from the resulting sample requires moderately high spatial
resolution and benefits from good U-V coverage.

Sensitivity, frequency coverage, RFI and angular resolution have long
been the rocks on which radio telescopes have foundered in their
attempts to use 21~cm absorption studies as a probe of structure
formation in the Universe.  All of these issues will be addressed in
the SKA and are briefly discussed below; the specifications are from
\cite{dayton04}.

{\bf 1. Sensitivity :} The specifications for the SKA sensitivity at
low frequencies are $A_{eff}/T_{sys} = 5000$ at 200~MHz and
$A_{eff}/T_{sys} = 20000$ between 0.5 and 5~GHz. These correspond to
$1 \sigma$ thermal noise values of 0.08~mJy and 0.013~mJy per $\sim
10$~km/s channel in one hour, at frequencies of 200~MHz ($z_{\rm abs}
\sim 6.1$) and 500~MHz ($z_{\rm abs} \sim 1.8$), respectively.

We will use an HI column density of $2 \times 10^{20}$~\scm ~(i.e. the
``classic'' definition of a DLA; \cite{wolfe86}) and a high spin
temperature of $5000$~K to estimate the background flux densities
toward which the SKA will be able to detect 21~cm absorption. Of
course, colder gas will be detected against weaker sources. Note that,
while WNM kinetic temperatures are likely to lie in the range $\sim
5000 - 10000$~K \cite{wolfire95}, Figure~2 of \cite{liszt01} shows
that these correspond to $\ts \lesssim 5000$~K, for two-phase media
and pressures $P/k \lesssim 6000$~cm$^{-3}$~K. We hence use $\ts =
5000$~K for the current estimates.

Even at its lowest sensitivity, at 200~MHz, equation~(\ref{eqn:21cm})
shows that 12~hour SKA integrations will be able to detect 21~cm
absorption at the $5~\sigma$ level toward 100~mJy background sources,
for the above values of $\NHI$ and $\ts$.  At lower absorption
redshifts, $z_{\rm abs} \lesssim 2$, where the SKA has a significantly
higher sensitivity, it will detect absorption toward far weaker
sources  (with flux densities $\sim 20$~mJy in the same integration
time), as well as detecting the absorption against the numerous
100 mJy sources in half hour integrations that are suitable for 
extensive surveys. Note that even deep SKA
integrations lasting 360~hours will only detect M$^\star_{\rm HI}$
galaxies
in 21~cm {\it emission} out to redshifts $z \sim 2.5$, and
0.1~M$^\star_{\rm HI}$ systems out to $z \sim 1$. 
(Galaxies at the knee of the $z\approx 0$ luminosity function have
M$^\star_{\rm HI}=10^{9.79}$M$_{\odot}$
for $H_o=$75~km~s$^{-1}$Mpc$^{-1}$\cite{zwaan03}.) Clearly, it will be
difficult to use 21~cm emission searches to trace galactic evolution,
if hierarchical merger models are indeed correct. And, since a high
signal-to-noise ratio is needed for kinematic studies, these will only
be possible out to $z \sim 1$, even for 360~hour integrations on
M$^\star_{\rm HI}$ galaxies.

{{\bf 2. Frequency coverage and RFI :} The SKA will have
frequency coverage from $\sim 100$~MHz to $\sim 25$~GHz and will thus
be sensitive to 21~cm absorption out to redshifts $z_{\rm abs}
\lesssim 13$. It is likely to be built on a site with low terrestrial
radio background is low, and it will also be one of the first radio
telescopes to incorporate modern RFI mitigation techniques
(e.g. \cite{briggs00b,fridman01}), crucial for observations at these
frequencies. These techniques will make it possible for the SKA to
reach theoretical noise levels in unprotected radio bands.}

{{\bf 3. Survey capabilities :} The SKA will have a wide-band correlator 
with the ability to handle $10^4$~channels over a large input bandwidth. 
In combination with its wide field of view (200~square degrees at 0.7~GHz)
large instantaneous bandwidth (2 bands, each with a full width equal to a 
fourth of the centre frequency of the observing band) and unequaled low 
frequency coverage, the SKA will be wonderfully suited for blind 21~cm 
absorption surveys, never before possible with an interferometer.}

{{\bf 4. Angular resolution :} The SKA will have an angular resolution
better than $({0.02/\nu_{\rm GHz} })''$, with excellent U-V
coverage. This corresponds to spatial resolutions of $\sim 350$~pc at
$z\sim 2$ and $\sim 600$~pc at $z \sim 6$. Remarkably, this is
somewhat better than the typical spatial resolution obtained in HI
emission studies of the kinematics of local galaxies
(e.g. \cite{swaters02}).}

\section{Blind SKA surveys}
\label{sec:surveys}

The sensitivity of a blind survey for redshifted absorption can be
quantified by the total redshift path surveyed and the sensitivity
function (i.e. the number of lines of sight at a given redshift with
detectable absorption; e.g.  \cite{storrie00}). In the case of the
SKA, the sensitivity function of a 21~cm absorption survey (out to $z
\sim 13$) is determined by the number of radio sources available as
background targets at a given redshift, with sufficiently high flux
density to show detectable 21~cm absorption.  We have seen in
Section~\ref{sec:ska} that twelve hours of SKA integration will be
sufficient to detect 21~cm absorption at the $5\sigma$ level in all
DLAs ($N_{HI}>2{\times}10^{20}$~cm$^{-2}$ and $T_s < 5000$~K) toward
100~mJy sources out to $z \sim 6$ and toward even weaker sources at
lower redshifts. We will hence use a flux cut-off of 100~mJy at the
redshifted 21~cm frequency to estimate the number of sources available
as background targets for a blind survey.

\begin{figure}
\epsfig{file=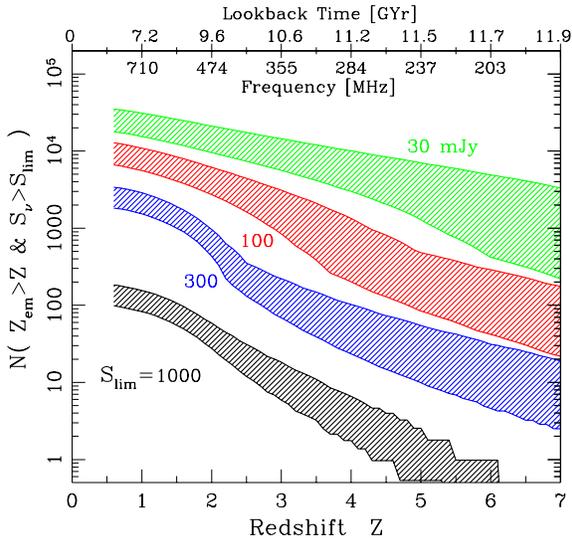,width=3.0in}
\caption{The number of background radio galaxies per $2\pi$ steradians
with redshift greater than $z$  and flux density greater than $S_{lim}$
at frequency $f=1420(1+z)^{-1}$MHz,
for $S_{lim}=30$, 100, 300 and 1000~mJy.  The bands indicate the range of 
uncertainty that spans the no-evolution and mild ($1 \sigma$) evolution
cases \cite{jarvis01}. 
}
\label{fig:numbers_gt_Z}
\end{figure}

The temporal evolution of the bright end of the radio luminosity
function is reasonably well-understood out to $z \sim 4$, based on low
frequency radio samples \cite{jarvis00,jarvis01,willott01}.  We have
used the results from \cite{jarvis01} to estimate the number of radio
galaxies available as absorption targets for the SKA. Note that radio
galaxies form ideal background sources, since they have sufficient
angular extent to probe galactic scales (one to tens of kpc) in a
foreground absorber. Fig~\ref{fig:numbers_gt_Z} shows the number of
radio sources above a given flux density that will be available for
absorption studies at different redshifts; this has been done for four
limiting flux densities, $S_{\lim} =$ 30, 100, 300 and 1000~mJy,
calculated at the redshifted 21~cm line frequency. The bands in the 
figure indicate the range of uncertainty spanning the
no-evolution and mild ($1~\sigma$) evolution cases described by
\cite{jarvis01}.  We restrict the following discussion to $z \lesssim
4$ to avoid speculative extrapolations in source number counts.

Fig.~\ref{fig:numbers_gt_Z} shows that there should be of order $10^3$
radio sources per hemisphere of the sky with $z > 4$, whose flux
densities at 284~MHz (corresponding to the HI line at $z=4$) exceed
100 mJy. The survey sensitivity function at $z \sim 4$ is thus $\sim
1000$, a factor of 20 larger than the combined sensitivity of all
current optical DLA surveys at this redshift (see Fig.~5 of
\cite{storrie00}).  The DLA statistic $n_{DLA}(z)=dN/dz$ quantifies the
expected number of interceptions of $\NHI>2{\times}10^{20}$cm$^{-2}$
absorbers that will occur per unit redshift. This interception rate
rises steadily to the range 0.3 to 0.4 for redshifts greater than 4
\cite{prochaska04}.
Note that the $z \sim 4$ sources probe absorbing
galaxies through 90\% of the age of the Universe. Further, a number of
30~mJy sources will show ``associated'' absorption at these redshifts,
which will be stronger due to the higher metallicity and dust content
near the nuclei of the radio galaxies and quasars.  This will allow
the measurement of redshifts of young radio galaxies (regardless of
their dusty cocoons), independent of optical spectroscopy.

Of course, the SKA sensitivity is at its lowest below $\sim 300$~MHz
\cite{dayton04}.  At $\sim 500$~MHz (i.e. $z \sim 1.85$, for the HI
line), it will be possible to detect 21~cm absorption in DLAs toward
20~mJy sources (see Section~\ref{sec:ska}).  The survey sensitivity
function at $z \sim 2$ is thus $ > 2 \times 10^4$, more than two
orders of magnitude larger than the combined value from present
optical surveys \cite{storrie00}.

Clearly, blind SKA 21~cm absorption surveys will allow the construction
of unparalleled samples of ``normal'' galaxies out to $z \gtrsim 4$. As mentioned 
earlier, these will be unbiased by dust extinction and thus well-suited to 
study the evolution of typical galaxies with redshift.  We note, finally, 
that observing time requirements are drastically alleviated by the large beam 
of the SKA ($\sim 200$~square degrees at 0.7~GHz; \cite{dayton04}); this allows 
spectra of a large number of sources to be taken simultaneously, at any given 
frequency setting.

In summary, the SKA will be able to carry out 21~cm absorption
measurements against thousands of distant radio sources out to
redshifts of four or greater, allowing detailed studies of galaxies
that intervene by chance along the line of sight.  These include the
kinematics and dynamical masses of the absorbers, their gaseous extent
and the structure of their ISM. In addition, blind 21~cm absorption
surveys will give rise to large unbiased samples of gas-rich
intervening galaxies, allowing us to determine the size and mass of
normal galaxies as a function of redshift and to thus directly test
theoretical models of structure formation. These surveys will also
provide large samples of associated absorbers, enabling studies of the
evolution in the environment of active galactic nuclei from high
redshifts to the present epoch.

\vskip 0.2in 
\noindent {\bf Acknowledgments} We are grateful to Sandhya Rao and Susan Ridgway 
for kindly providing us with the damped Lyman-$\alpha$ profile toward 
QSO~0738+313 and a fully processed HST image of 3C196, respectively. NK thanks 
Jayaram N Chengalur for permission to present unpublished results on 3C196 and 
PKS~1229--021.
\bibliographystyle{h-elsevier2} 
\bibliography{ms}
\expandafter\ifx\csname natexlab\endcsname\relax\def\natexlab#1{#1}\fi

\end{document}